\documentclass[prb,twocolumn,showpacs,preprintnumbers,superscriptaddress,amsmath,amssymb,10pt]{revtex4-1}

\usepackage{color}
\definecolor{rot}{rgb}{1,0,0}

\usepackage[T1]{fontenc}
\usepackage[pdftex]{graphicx}
\usepackage{dcolumn}
\usepackage{bm}

\usepackage{mathcomp}

\DeclareGraphicsExtensions{.pdf}
\usepackage[justification=raggedright,singlelinecheck=false]{caption}

\begin{document}

\title{Indirect doping effects from impurities in MoS$_2$/BN heterostructures}
\author{Roland Gillen}\email{r.gillen@tu-berlin.de}
\affiliation{Institut für Festkörperphysik, Technische Universität Berlin, Hardenbergstr. 36, 10623 Berlin, Germany}
\affiliation{Center for Advanced Photonics and Electronics, Department of Engineering, University of Cambridge, 9 JJ Thomson Avenue, Cambridge CB3 0FA, United Kingdom}
\author{John Robertson}
\affiliation{Center for Advanced Photonics and Electronics, Department of Engineering, University of Cambridge, 9 JJ Thomson Avenue, Cambridge CB3 0FA, United Kingdom}
\author{Janina Maultzsch}
\affiliation{Institut für Festkörperphysik, Technische Universität Berlin, Hardenbergstr. 36, 10623 Berlin, Germany}

\date{\today}

\begin{abstract}
We performed density functional theory calculations on heterostructures of single layers of hexagonal BN and MoS$_2$ to assess the effect of doping in the BN sheet and of interstitial Na atoms on the electronic properties of the adjacent MoS$_2$ layer. Our calculations predict that $n$-doping of the boron nitride subsystem by oxygen, carbon and sulfur impurities causes noticeable charge transfer into the conduction band of the MoS$_2$ sheet, while $p$-doping by beryllium and carbon leaves the molybdenum disulphide layer largely unaffected. Intercalated sodium atoms lead to a significant increase of the interlayer distance in the heterostructure and to a metallic ground state of the MoS$_2$ subsystem. The presence of such $n$-dopants leads to a distinct change of valence band and conduction band offsets, suggesting that doped BN remains a suitable substrate and gate material for applications of $n$-type MoS$_2$.
\end{abstract}


\maketitle

\section{Introduction}
Semiconductor materials of the family of dichalcogenides, such as molybdenum disulfide (MoS$_2$), attracted significant interest over the last years by virtue of their potential for future applications in electronics~\cite{Radisavljevic-2011,kim-2012,lembke-2012, sangwan-2013} and optoelectronics\cite{wang-2012, lee-2012, zhang-2014}. Similar to graphite, these materials consist of stacked quasi-twodimensional atomic layers that are physically independent to a large extent and can be separated easily. 
Particularly interesting from the application point-of-view is the combination of two-dimensional (2D) dichalcogenides with other two- or low-dimensional materials, such as graphene\cite{roy-2013, zhang-2014}, boron nitride or other dichalcogenides\cite{terrones-2013}. This in principle opens the door to a whole new class of artificial materials with designed electronic, optical and mechanical properties\cite{novoselov-2012}. Due to its large band gap and surface smoothness, boron nitride might here be a good candidate as a substrate, tunnel layer or gate material, similar to its proposed use in graphene devices\cite{dean-2010}. 

Sufficient understanding of the effects of interfaces and heterostructures with other materials on the electronic and optical properties of quasi-2D dichalcogenides is an important prerequisite for design and realization of such applications. This involves, for instance, the interaction of 2D semiconductors with metal contacts or insulators, which could be used as dielectrica or substrate materials. As has been shown before\cite{Radisavljevic-2011}, dielectrica can have a positive effect on the electron  mobility in the single dichalcogenide layers in transistors by damping phonon-induced scattering and by screening of charged impurites\cite{kaasbjerg-2012}. On the other hand, defects in the interface region between substrate/dielectric and the quasi-2D layer have an additional effect on the physical properties of the system and could, for instance, cause intentional or unintentional doping of the semiconductor layer.

However, despite the increasing number of studies of the fundamental physical properties of quasi-2D dichalcogenides\cite{mak-2010,splendiani-2010,molina-sanchez-2011,korn-2011,rice-2013,sun-raman,qiu-2013,komsa-2013,chen-2013,song-2013}, systematic theoretical investigations on interfaces and heterostructures involving MoS$_2$ are scarce. 
Recently, calculations of the electronic band structures of molybdenum and tungsten dichalcogenides and Moir\'{e} patterns in the electronic structure of bilayer MoS$_2$/MoSe$_2$ systems have been reported\cite{kang-2013-bandoffsets,kang-2013-MoS2MoSe2}, while Komsa \textit{et al.}\cite{komsa-2013} presented absorption spectra of MoS$_2$/WS$_2$ bilayers from quasiparticle calculations and bandstructures for compounds of MoS$_2$ with graphene and boron nitride. Dolui \textit{et al.}, on the other hand, investigated defects as possible causes of the experimentally found intrinsic $n$- and $p$-doping of MoS$_2$ on SiO$_2$ substrates\cite{dolui-2013}. Investigations of such defects are of particular importance in two-dimensional systems, where weak dielectric screening of charged internal and external defects and breaking of the periodicity of the electrostatic potential have a strong effect on the electronic and transport properties. External defects might in principle also be exploited for intentional indirect doping of thin semiconductor layers by suitable doping of the underlying substrate.  
Yet, such simulations for MoS$_2$ on other substrates, or for heterostructures of MoS$_2$ on BN or MoS$_2$ on graphene do not exist to date.

The aim of this paper is thus an assessment of the effect of defects in heterostructures of monolayer MoS$_2$ with layered boron nitride using a density functional theory (DFT) approach. We focus on intentionally and unintentionally introduced defects in the BN subsystem, \textit{e.g.}, vacancies, $p$-type and $n$-type impurities, and intercalants in the interstitial region. Our results show that (i) $n$-dopants incorporated into the BN sheet lead to $n$-doping of the MoS$_2$ due to charge transfer over the interstitial region, while (ii) $p$-doping by C and Be atoms shows a considerably weaker effect on the electronic properties of MoS$_2$. (iii) Intercalated Na atoms lead to a significantly increased BN-MoS$_2$ layer distance and also to charge transfer to MoS$_2$ due to Fermi level pinning. We further show that (iv) the hybrid functional HSE12\cite{hse12} can successfully describe the electronic band gaps of both BN and MoS$_2$ on a similar level as the \textit{GW} approximation.

\section{Method}
\begin{figure}[tbh]
\centering
\begin{minipage}{\columnwidth}
\includegraphics*[width=0.85\textwidth]{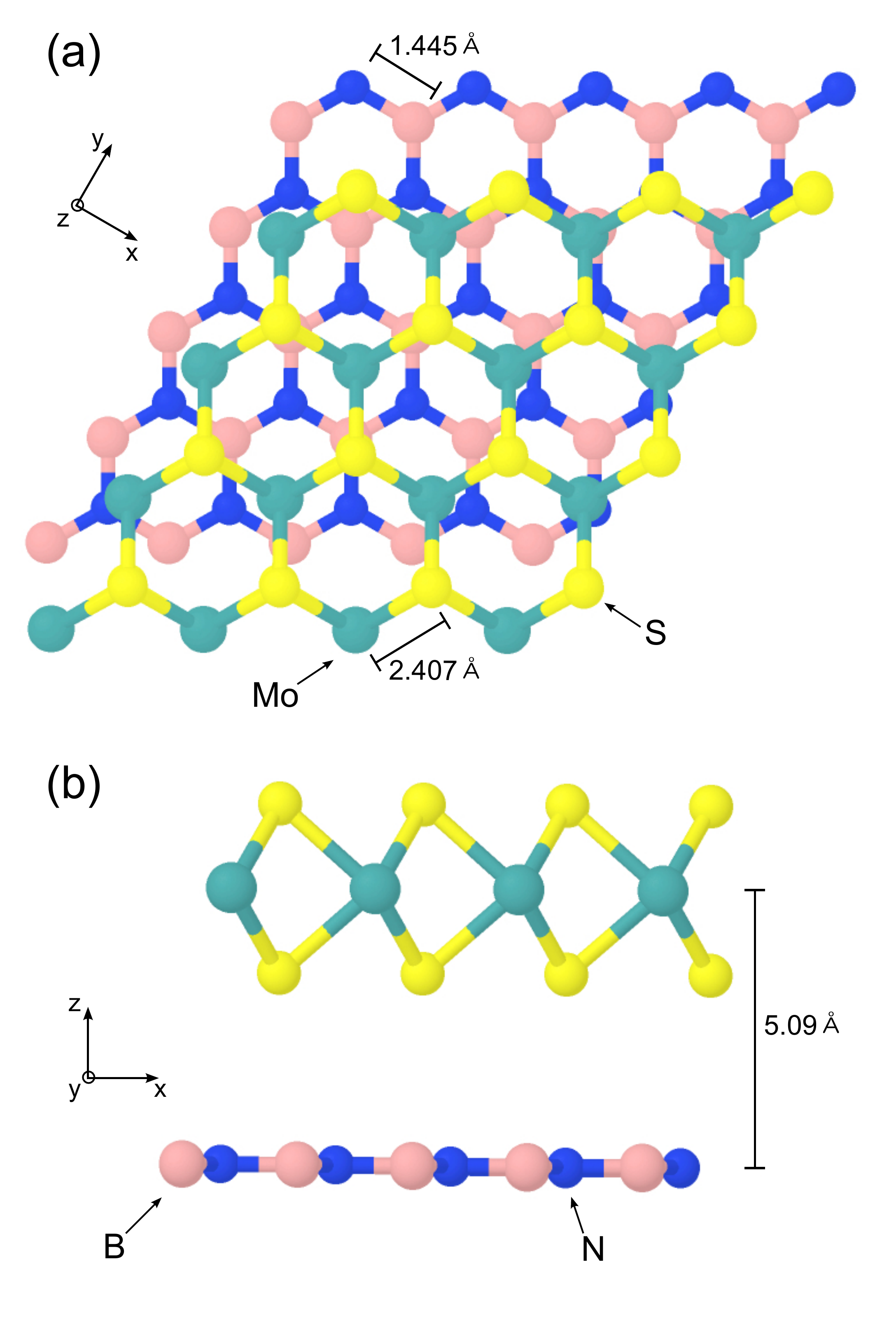}
\end{minipage}
\caption{\label{fig:pure-geom} (Color online) Top and side view of the unit cell of a heterostructure of MoS$_2$ on top of a BN layer used in the calculations.
}
\end{figure}
The presented results have been calculated by the use of the reparametrized Perdew-Becke-Ernzerhof exchange correlation functional for solids (PBEsol)\cite{pbesol} as implemented in the planewave code CASTEP\cite{castep}. We modeled the MoS$_2$/BN heterostructures by putting a 4x4x1 supercell of single-layer MoS$_2$ (lattice constant a$_{\mbox{\tiny MoS2}}$=3.17\,\AA) on top of a 5x5 supercell of BN (a$_{\mbox{\tiny BN}}$=2.5\,\AA), which limited the lattice mismatch between the two layers to ~1.4\% and resulted in a simulation cell containing 98 atoms in the defect-free case. We varied the initial relative positions of the MoS$_2$ and BN sheets but only found a negligible dependence of the total energy on the stacking within the margins of error. We then introduced one defect per supercell by replacing a single boron or nitrogen atom by the dopant of interest, adding an atom in the interfacial vacuum region between the layers or removing single atoms to create vacancies, and then optimized the atomic positions until the residual interatomic forces were below 0.01\,eV/\,\AA, while keeping the cell parameters fixed at the value of 12.5\,\AA\space of the pure BN supercell. The dimensions of the supercell were sufficiently large to contain all geometric relaxations of the atomic shells surrounding the studied defects.

To minimize interactions between periodic images due to 3D boundary conditions, we introduced a vacuum layer such that the distance between periodic images was at least 25\,\AA. 
We modeled the interaction of the valence electrons with the pseudoatomic cores of all the atomic species present in our studied structures by normconserving pseudopotentials generated with the OPIUM\cite{opium1,*opium2} package with a cutoff energy of 800 eV, explicitly including the semi-core Mo $4d$ electrons in the calculations. The integrals in reciprocal space were performed on a discrete 2x2x1 Monkhorst-Pack \textit{k}-point grid for which the total energy of the pure heterostructure was converged to below a value of 0.01\,eV. Following a ground state calculation, the energies in the partial density-of-states (PDOS) plots were derived from band structure calculations on a 10x10x1 \textit{k}-point grid and broadened by Gaussians of width 0.1\,eV.

\section{Results and Discussion}
\subsection{Pure MoS$_2$/BN heterostructure}\label{sec:pure}
\begin{figure}[tbh]
\centering
\begin{minipage}{0.98\columnwidth}
\includegraphics*[width=\textwidth]{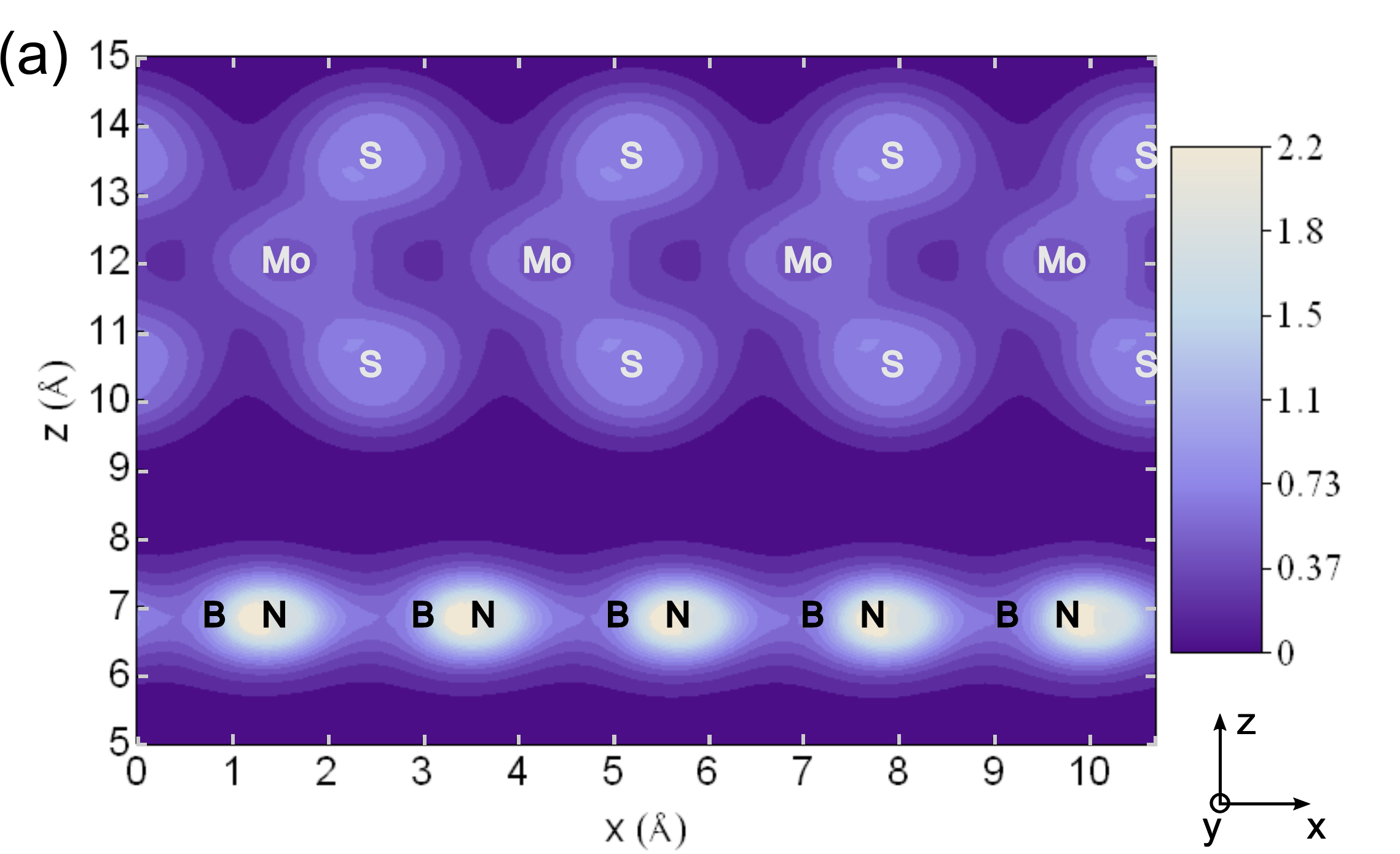}
\end{minipage}
\\
\begin{minipage}{0.98\columnwidth}
\includegraphics*[width=\textwidth]{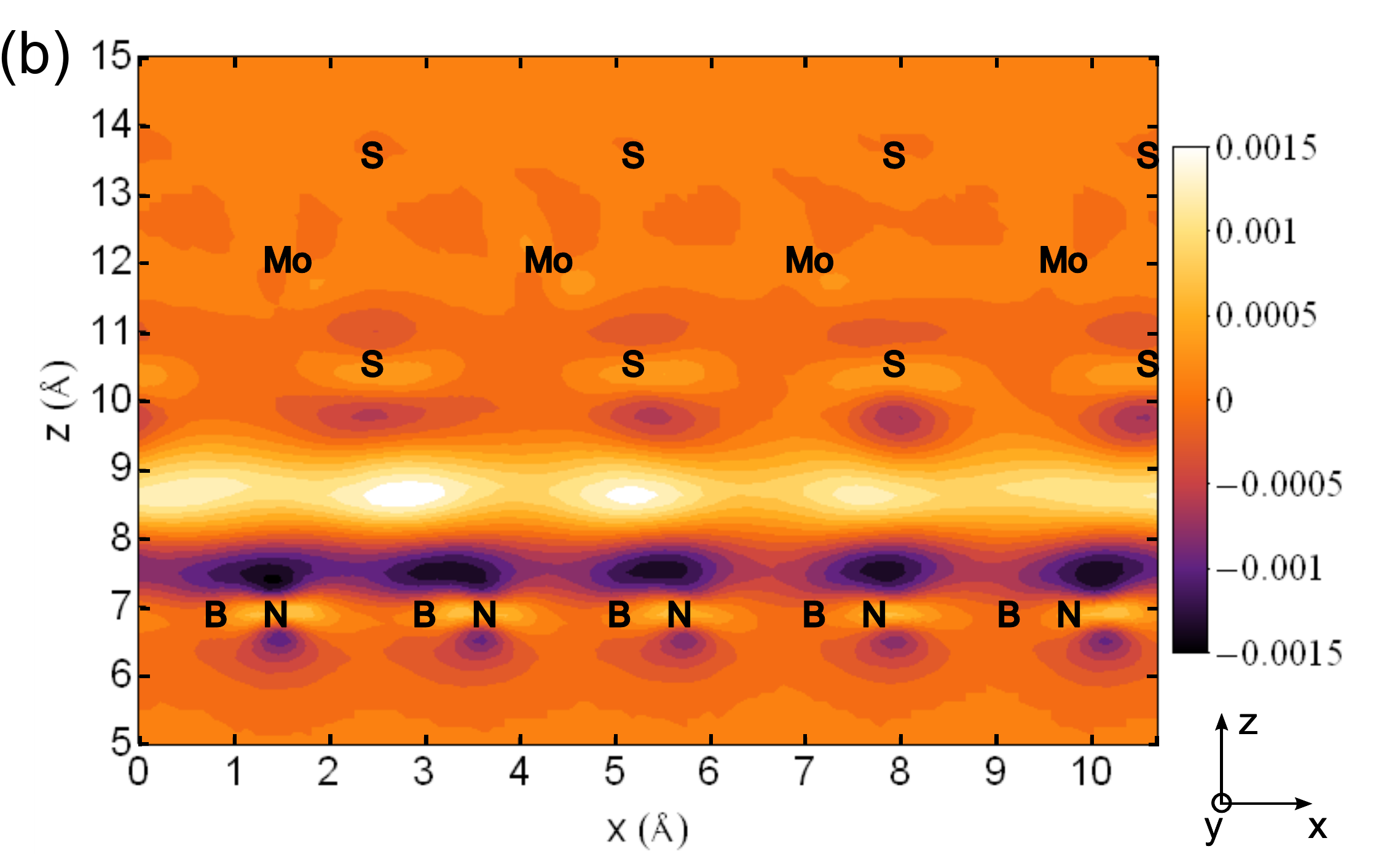}
\end{minipage}
\\
\begin{minipage}{0.98\columnwidth}
\includegraphics*[width=0.7\textwidth]{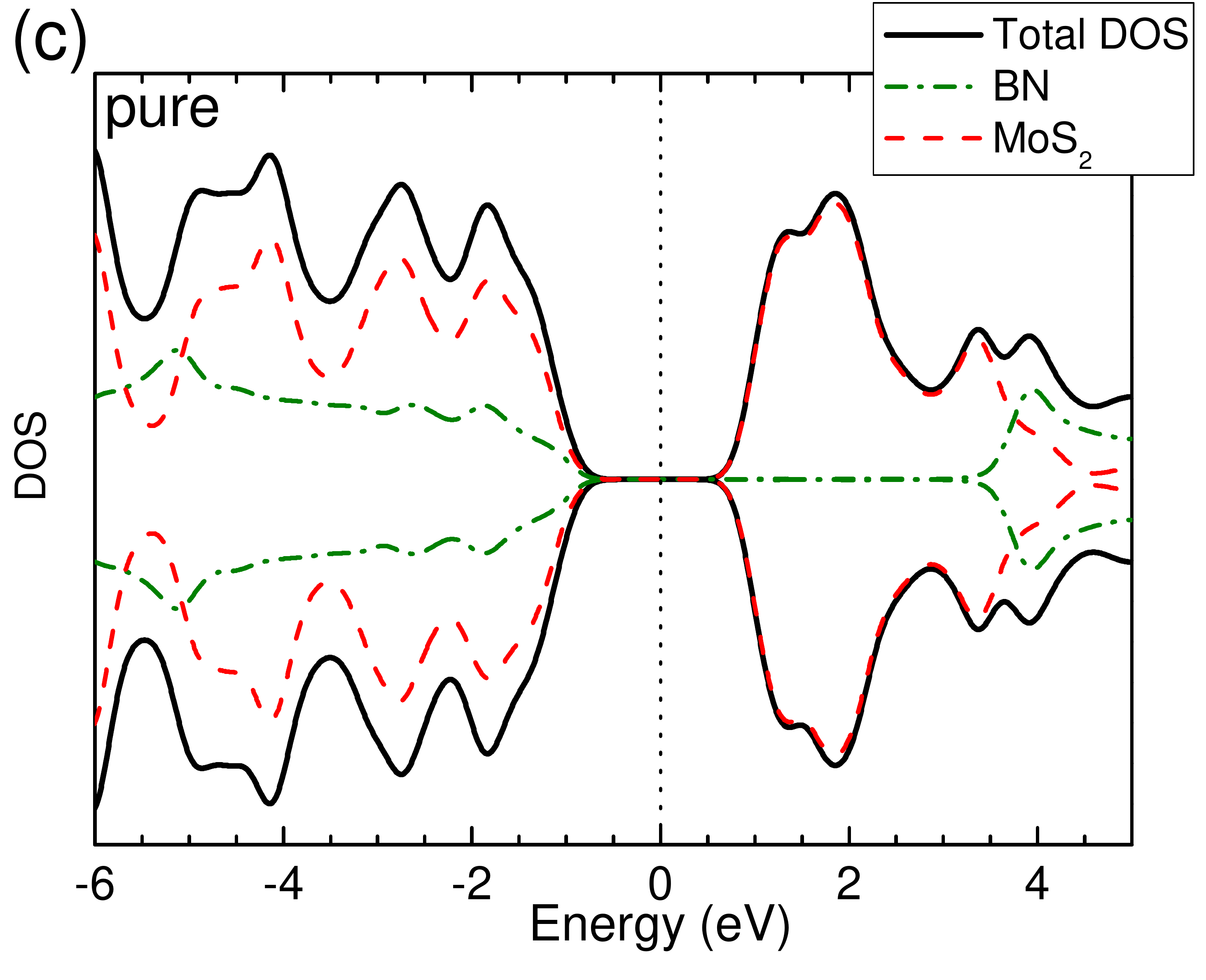}
\includegraphics*[width=0.28\textwidth]{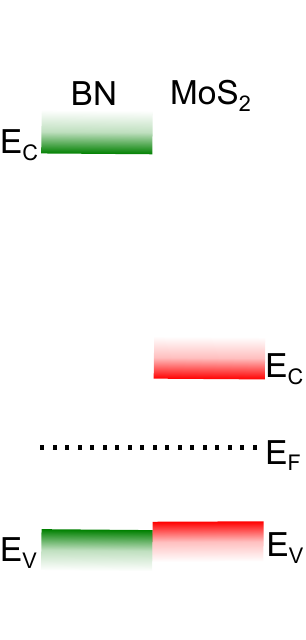}
\end{minipage}
\caption{\label{fig:pure} (Color online) Pristine bilayer MoS$_2$/BN heterostructure: (a) The electron density (in electrons/\AA$^3$) averaged over the $y$-axis. (b) The corresponding charge density difference (also in electrons/\AA$^3$) as calculated from substracting the charge densities of the isolated BN and MoS$_2$ layers from that in (a). (c) Spin-resolved partial density of states with the zero-of-energy placed at the Fermi energy (dotted black line). The right part shows a schematic figure of the aligned band edges in the heterostructure.
}
\end{figure}
\begin{figure}[tbh]
\centering
\begin{minipage}{\columnwidth}
\includegraphics*[width=0.49\textwidth]{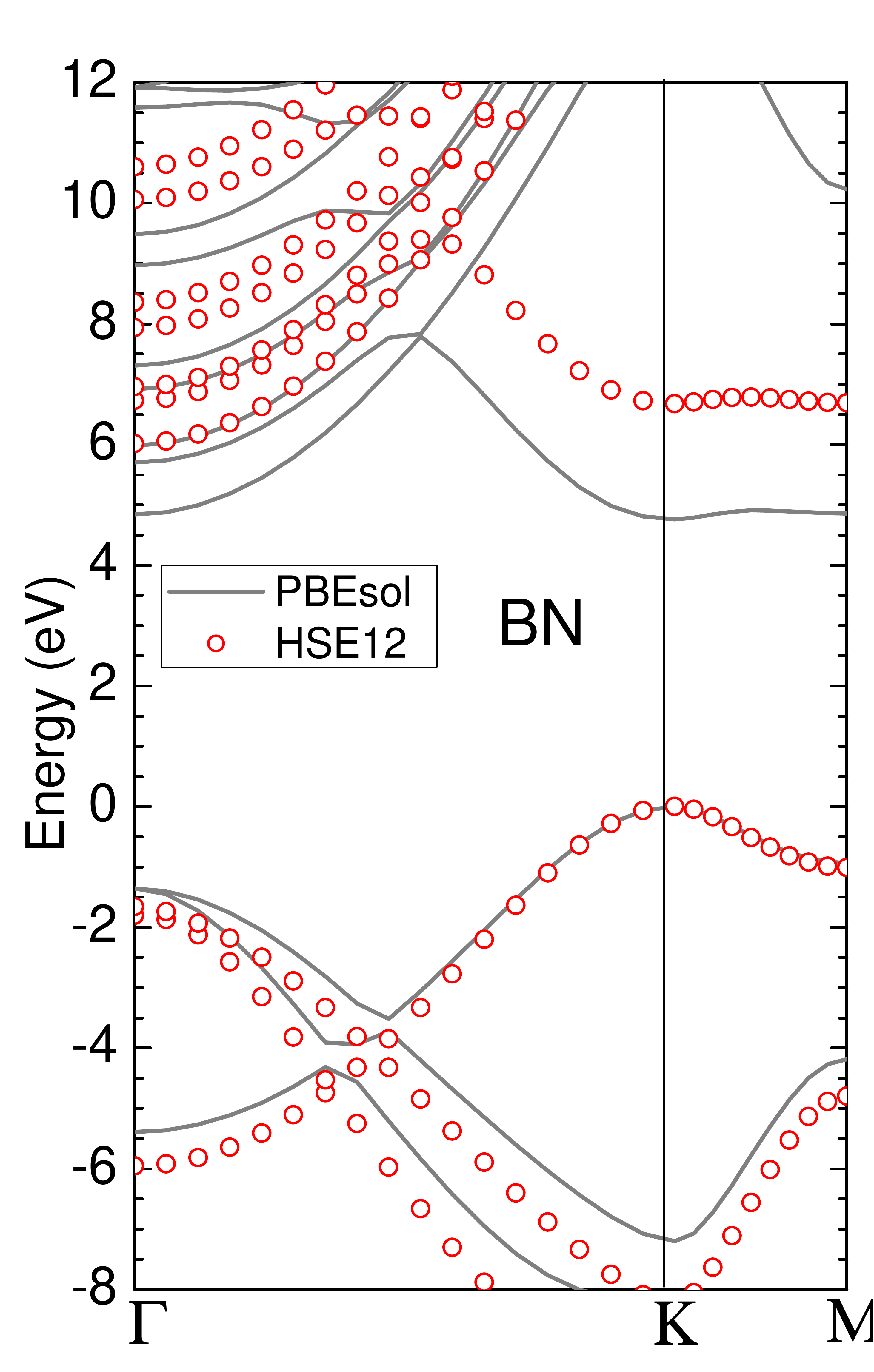}
\includegraphics*[width=0.49\textwidth]{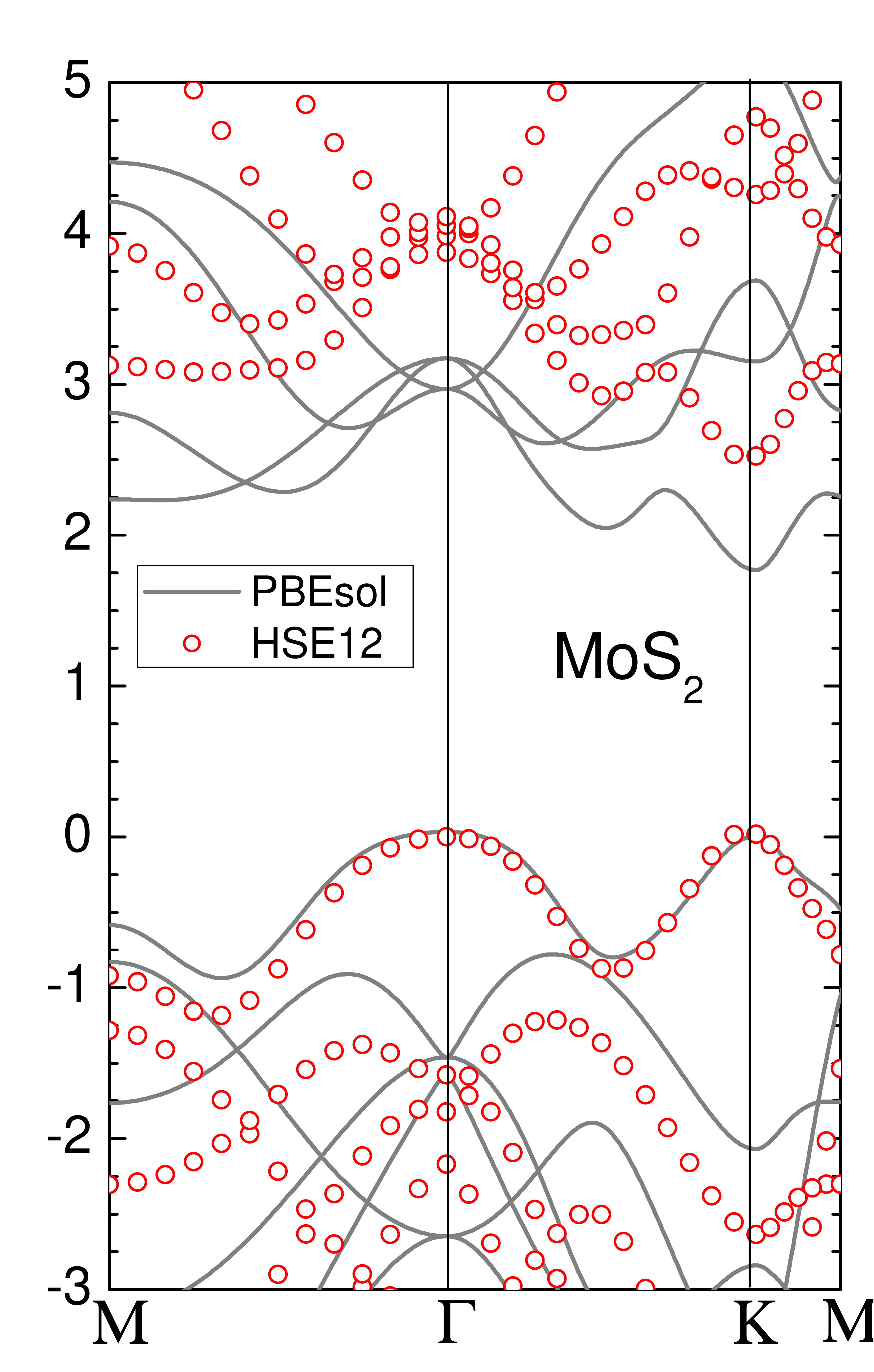}
\end{minipage}
\caption{\label{fig:BN-MoS2-bands} (Color online) Electronic band structures of BN and MoS$_2$ without spin-orbit coupling from calculations using the PBEsol (black lines) and the HSE12 (red dots) functionals. For purposes of comparison, the bands are shifted such that the valence band maxima coincide with the zero-of-energy.
}
\end{figure}
We start with the electronic structure and geometry of the defect-free heterostructure of a MoS$_2$ layer stacked on a BN sheet, which we will use for comparison with the defective structures through the course of the paper. We find an average distance of the Mo atoms from the BN sheet of 5.09\,\AA\space on the level of the PBEsol functional, together with B-N and Mo-S bond lengths of 1.445\,\AA\space and 2.407\,\AA, respectively, see Fig.~\ref{fig:pure-geom}. 
Explicit inclusion of van-der-Waals interactions through the semi-empirical Grimme 2006 functional\cite{g06} led to a decreased layer distance of 4.88\,\AA, but this did not affect the electronic properties relevant for the results of this paper. 

Altogether, the MoS$_2$ and BN layers are weakly bound together. Firstly, this is indicated by the fact that in-plane translations of the MoS$_2$ layer relative to the BN plane had a negligible effect on the reported results. Another indicator is the weak change of electron density distribution from stacking of the two layers. Figure~\ref{fig:pure}~(a) shows a plot of the electron density of the heterostructure projected onto the $x$-$z$ plane, obtained by averaging the calculated charge density over the $y$-axis. As expected, the projected charge density is the highest within the BN and MoS$_2$, particularly in the bonding region between boron and nitrogen atoms, where the projection gives rise to bright areas of high density in the plot.
The density in the interstitial region is significantly lower and close to zero. This is also reflected in the Mulliken population of the four species in the heterostructure. The values of $+0.82$\,\textit{e} (\textit{e} being the elementary charge) and $-0.83$\,\textit{e} for nitrogen and boron, respectively, $-0.05$\,\textit{e} for sulfur and $+0.13$\,\textit{e} for molybdenum are virtually unchanged compared to the isolated subsystems. This suggests the absence of noticeable charge transfer between the layers in the MoS$_2$/BN compound structure. In order to take a closer look at more subtle changes in electronic charge distribution in the heterostructure, we show in Fig.~\ref{fig:pure}~(b) the difference between the $y$-averaged density distribution shown in Fig.~\ref{fig:pure}~(a) and the ones from the isolated BN and MoS$_2$ layers. The plot reveals a small accumulation of charge in the interstitial region, smaller than the average density in the BN layer by a factor on the order of 10$^3$-10$^4$. This mainly stems from charge transfer from the BN layer and sulfur atoms in the Mo$S_2$ layer. The small magnitude of accumulated charge suggests only weak covalent character of the interlayer interaction in this system.

In accordance with previously reported results from Ref.~\onlinecite{komsa-2013}, we find the energies of the valence band maxima of BN and MoS$_2$ very close to each other. They appear to be almost aligned in the partial density of states (PDOS) of the compound system in Fig.~\ref{fig:pure}~(c), rendering MoS$_2$/BN a type-II heterostructure in our calculations. Correspondingly, the valence band offset of the two subsystems is close to zero, while the conduction band offset is given by the difference in the band gap sizes of BN and MoS$_2$. From our PBEsol calculations (neglecting spin-orbit interaction), we obtain an electronic band gap value of 1.75\,eV. This value is in good agreement with the experimental value of the \textit{optical} (excitonic) gap of around 1.8\,eV, which can be attributed to cancellation of the well-known underestimation of electronic band gap values, due to approximated electron-electron interactions, and missing electron-hole interactions in local DFT exchange-correlation functionals. The experimental value of the electronic band gap, on the other hand, is still a matter of debate. We thus also calculated the bandstructure of MoS$_2$ using the hybrid functional HSE12\cite{hse12} and found a higher band gap of 2.5\,eV, which is between the value found in a recently published study using the HSE06 functional\cite{dameng-2013} (2.3\,eV) and the quasi-particle gap from partially self-consistent \textit{GW}$_0$ calculations\cite{shi-2013} (2.66\,eV). The corresponding bandstructure is shown in Fig.~\ref{fig:BN-MoS2-bands}. We note that our HSE12 value is in good agreement with the predicted exciton binding energy of $\sim$0.6-0.9\,eV found from solving the Bethe-Salpether Equation\cite{qiu-2013,shi-2013}, which would lead to an optical gap of 1.6-1.9\,eV from our calculations.

Similarly, PBEsol significantly underestimates the electronic band gap of BN of about 6\,eV\cite{evans-2008}, predicting a value of 4.7\,eV. The non-local exchange included in the HSE12 functional successfully remedies this underprediction and opens the band gap to a value of 6.05\,eV, which is close to the experimental value and an improvement on the reported band gap value of 5.7\,eV from recent HSE06 calculations\cite{park-2012}. From that, we obtain a conduction band offset of about 2.95 eV for calculations based on the PBEsol functional and a higher value of 3.55 eV in case of the HSE12 functional. We thus conclude from our calculations that BN could act as a substrate/buffer layer for electron conduction but would be almost transparent for holes. An interesting question in this respect is whether interlayer interactions between the MoS$_2$ and BN subsystems are weak enough such that explicit consideration of spin-orbit interactions in MoS$_2$ in the calculations pushes the maximum at the $K$-point above that of the $\Gamma$-point, thus rendering the deposited MoS$_2$ layer a direct semiconductor like in isolated single-layers. Due to the significant computational ressources needed to properly address spin-orbit interactions for systems of this size, we will leave this investigation until a later point in time.

\subsection{B and N vacancies}
\begin{figure*}[tbh]
\centering
\begin{minipage}{\textwidth}
\includegraphics*[width=0.95\textwidth]{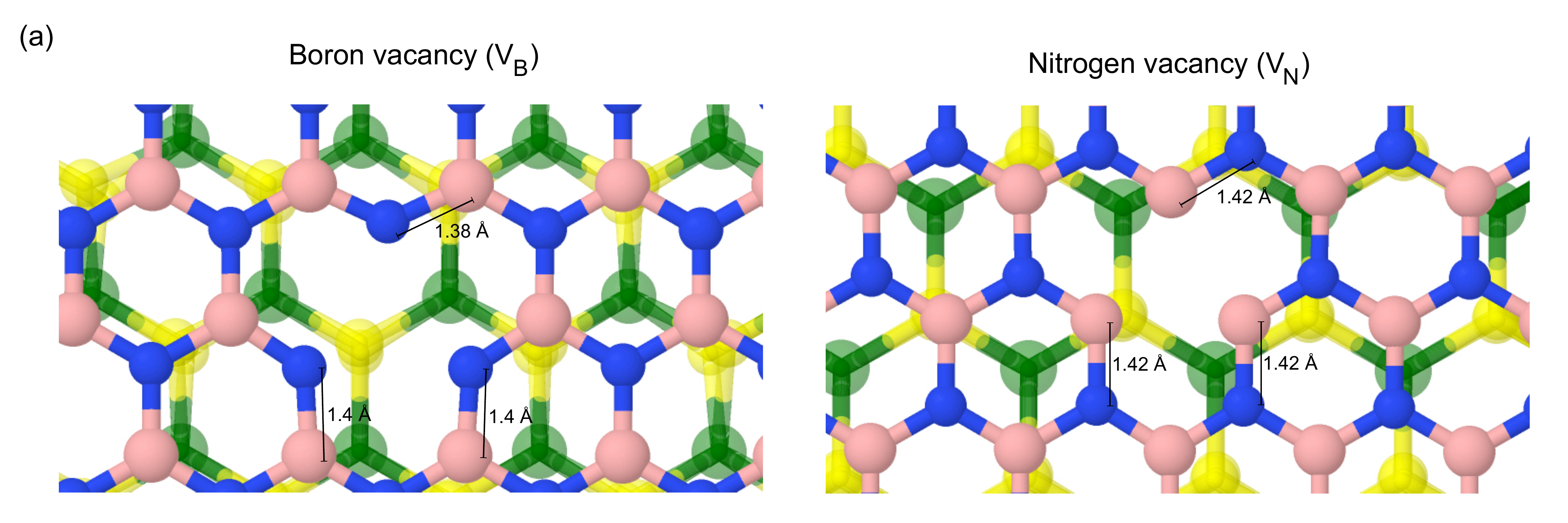}
\end{minipage}
\\
\begin{minipage}{0.98\columnwidth}
\includegraphics*[width=0.7\columnwidth]{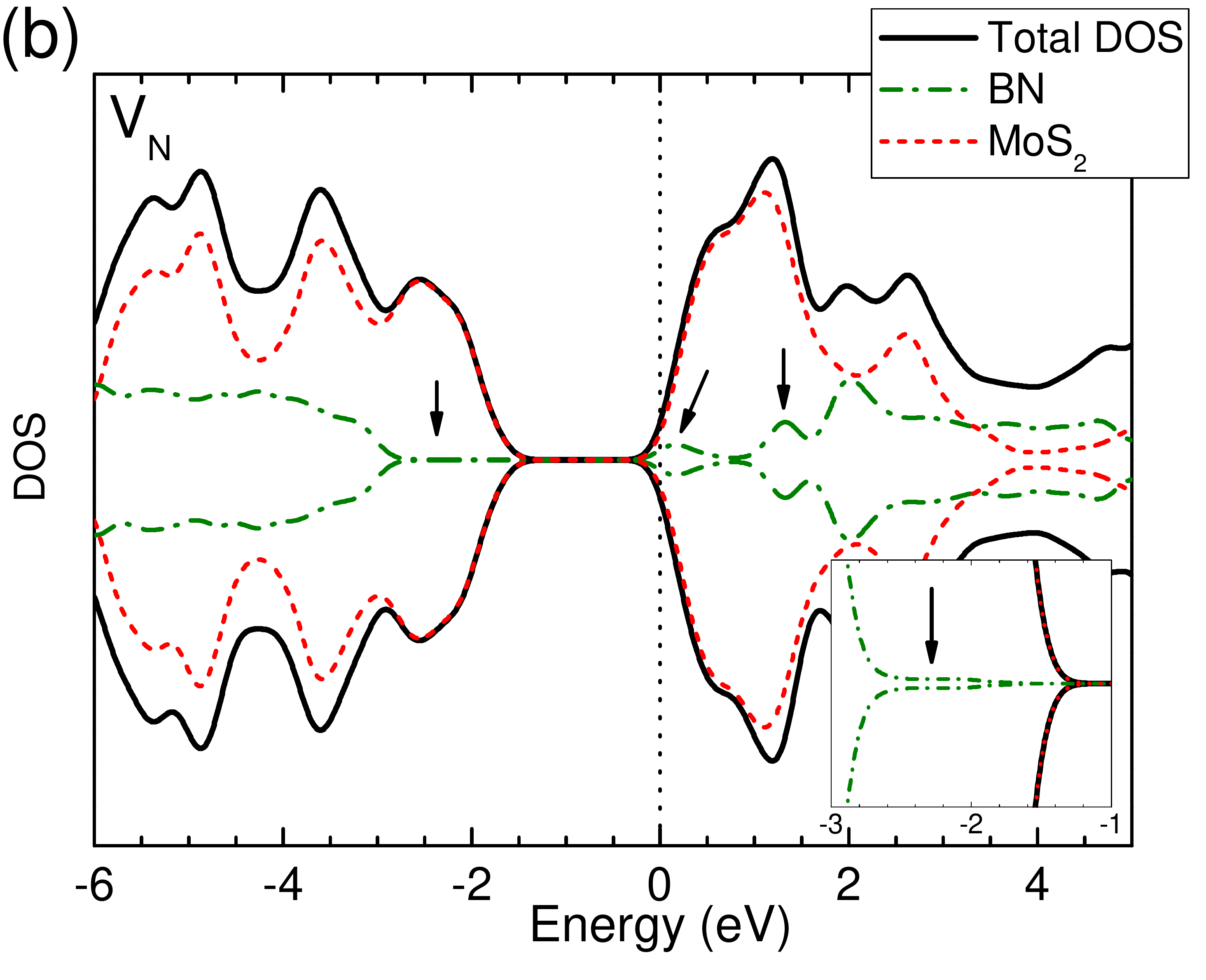}
\includegraphics*[width=0.28\columnwidth]{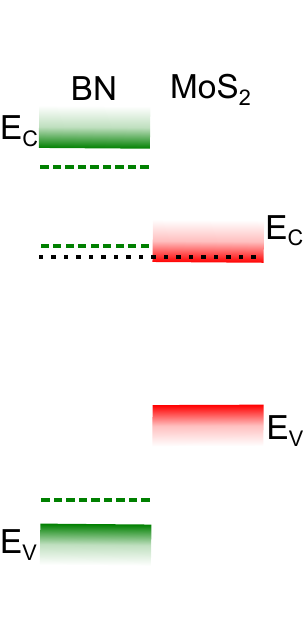}
\end{minipage}
\quad
\begin{minipage}{0.98\columnwidth}
\includegraphics*[width=0.7\textwidth]{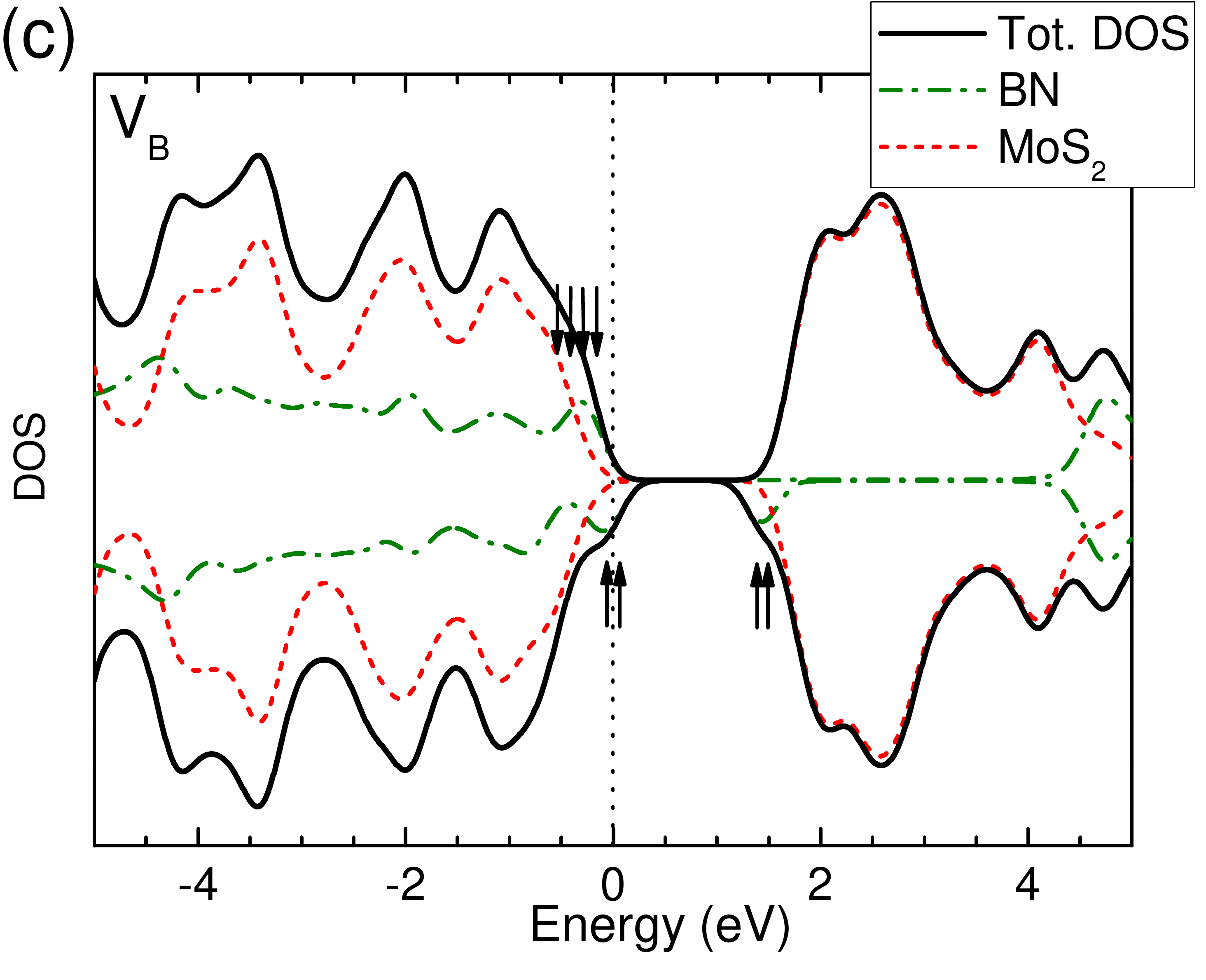}
\includegraphics*[width=0.28\textwidth]{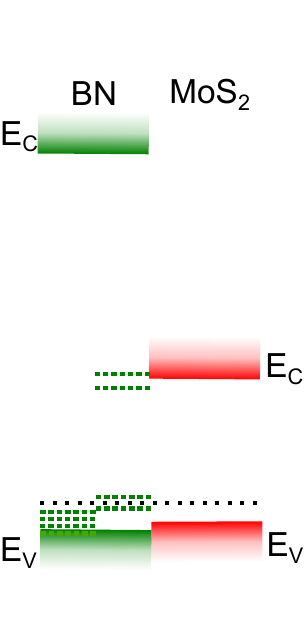}
\end{minipage}
\caption{\label{fig:vacancies} (Color online) (a) Defect geometries around  nitrogen ($V_N$) and boron ($V_B$) vacancies in the BN subsystem of a MoS$_2$/BN heterostructure. (b) and (c) show the corresponding spin-resolved partial density of states, where the zero-of-energy corresponds to the Fermi energy of the system (dotted lines). The black arrows in the DOS plot and the dashed (green) lines in the band scheme denote the energies of the defect levels induced by the vacancies.
}
\end{figure*}
An obvious choice for defects are native vacancies in the crystal lattice, which are expected to naturally occur during crystal growth but also can be induced intentionally by ion irradiation. The native vacancies in single-layer MoS$_2$ have recently been theoretically studied by Liu \emph{et al.}\cite{dameng-2013}, who found a significant effect of sulfur vacancies on the electronic properties of MoS$_2$ due to Fermi level pinning. In contrast, we focus in this work on more indirect effects from boron and nitrogen vacancies in the BN layer on the MoS$_2$ subsystem. We found that boron vacancies increase the distance of the Mo atoms to the BN layer by about 1\% to a value of 5.13\,\AA, while nitrogen vacancies do not significantly alter the layer distance. As was recently predicted to occur in the pure BN layer\cite{attaccalite-2011}, a small Jahn-Teller effect breaks the threefold symmetry of the nitrogen atoms surrounding the boron vacancy, see Fig.~\ref{fig:vacancies}~(a). Here, the B-N bond lengths around the defect decrease by 4\% to a value of 1.38\,\AA\space for bonds involving the symmetry-breaking nitrogen atom, and by 3\% to a value of 1.4\,\AA\space otherwise. The relaxation in case of the nitrogen vacancy is much smaller, the boron atoms undergo a breathing-like relaxation of about 1\%, down to a value of 1.42\,\AA.

Removing a nitrogen atom from the BN sheet introduces several defect states into the band structure, which correspond to four dangling bonds around the vacancy occupied by three electrons. The density-of-states plot in Fig.~\ref{fig:vacancies}~(b) reveals a fully occupied defect level slightly above the valence band top of the BN subsystem, with a wavefunction that is localized at the site of the vacancy and is composed of equal contributions from the three adjacent boron atoms. Two additional defect states, partially occupied with the remaining electron, appear closely below the conduction band minimum of BN and correspond to wavefunctions that are strongly localized at the boron atoms neighboring the vacancy. In the heterostructure, the energies of the partially occupied defect states are above the conduction band edge of MoS$_2$. A portion of the electronic charge in the defect levels is transferred from the BN layer into the energetically more favourable antibonding Mo and S states on the bottom of the MoS$_2$ layer, causing effective $n$-type doping conditions in the MoS$_2$ layer in the neutral charge state of the vacancy and a corresponding considerable realignment of the electronic bands compared to the defect-free case.

The boron vacancy in the neutral charge state, on the other hand, introduces eight spin-polarized single-electron defect levels closely above the BN valence band maximum, which are occupied by five electrons. The PDOS plot in Fig.~\ref{fig:vacancies}~(c) suggests that the four up-spin states are fully occupied by electrons, while the down-spin states are not or partially occupied. We thus obtain a spin-polarized ground state with a magnetic moment of 2.4$\mu_B$ ($\mu_B$ being the Bohr magneton), where the spin polarization is fully contained in the BN layer, mainly being localized at the nitrogen atoms surrounding the vacancy.

 The electronic structure of the MoS$_2$ layer is not directly affected in the zero temperature conditions of our calculations, as the unoccupied defect levels are still slightly above the valence band maximum of MoS$_2$. Still, it is likely that finite temperatures would allow for charge spilling from the sulfur atoms through the vacuum layer to the unoccupied acceptor states from the vacancy to some extent, leading to $p$-doping of the MoS$_2$.

\subsection{$n$-doping of BN}
\begin{figure}[tbh]
\centering
\begin{minipage}{0.98\columnwidth}
\includegraphics*[width=0.7\textwidth]{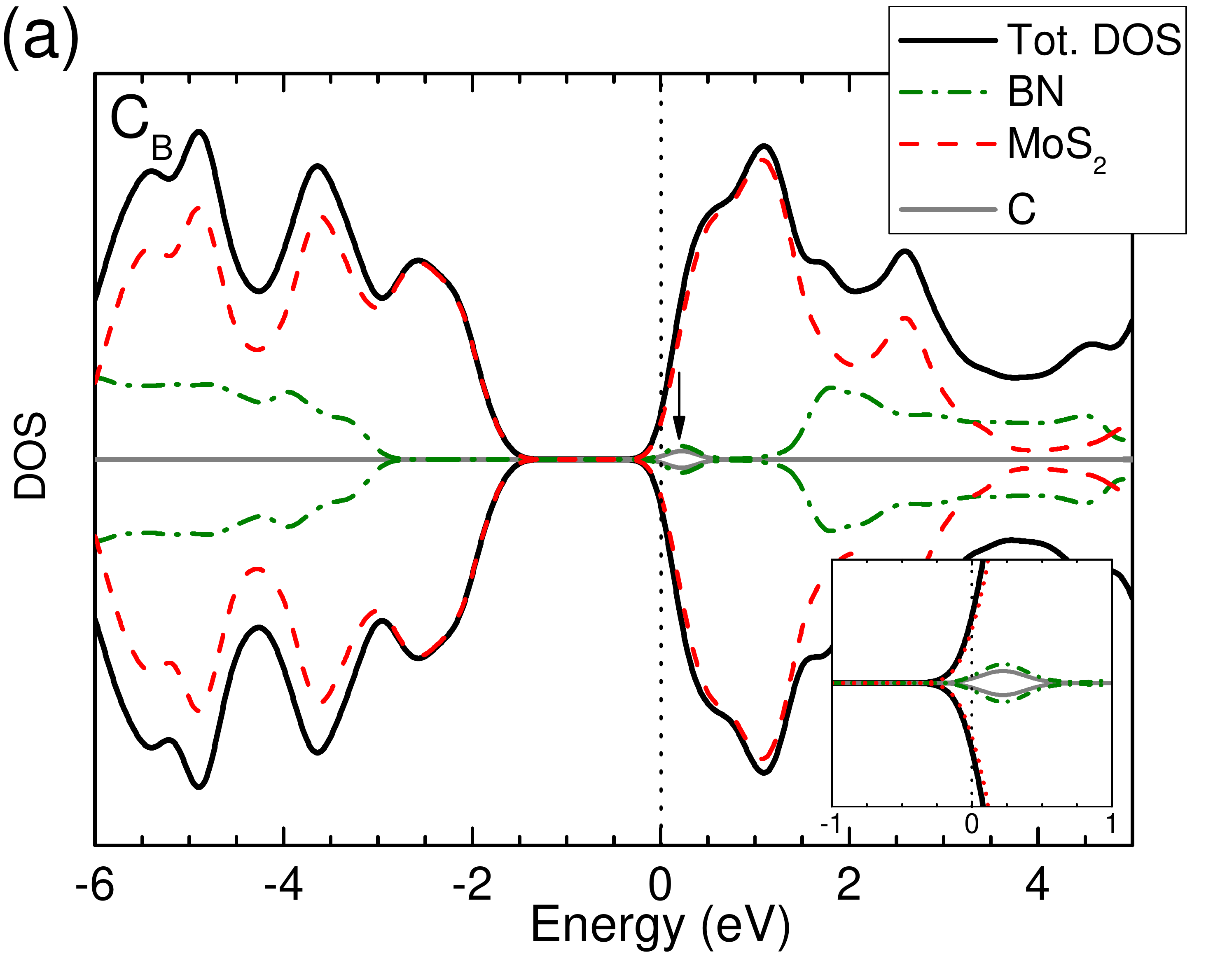}
\includegraphics*[width=0.28\textwidth]{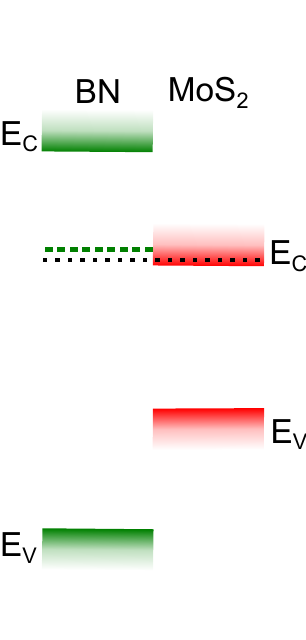}
\end{minipage}
\begin{minipage}{0.98\columnwidth}
\includegraphics*[width=0.7\textwidth]{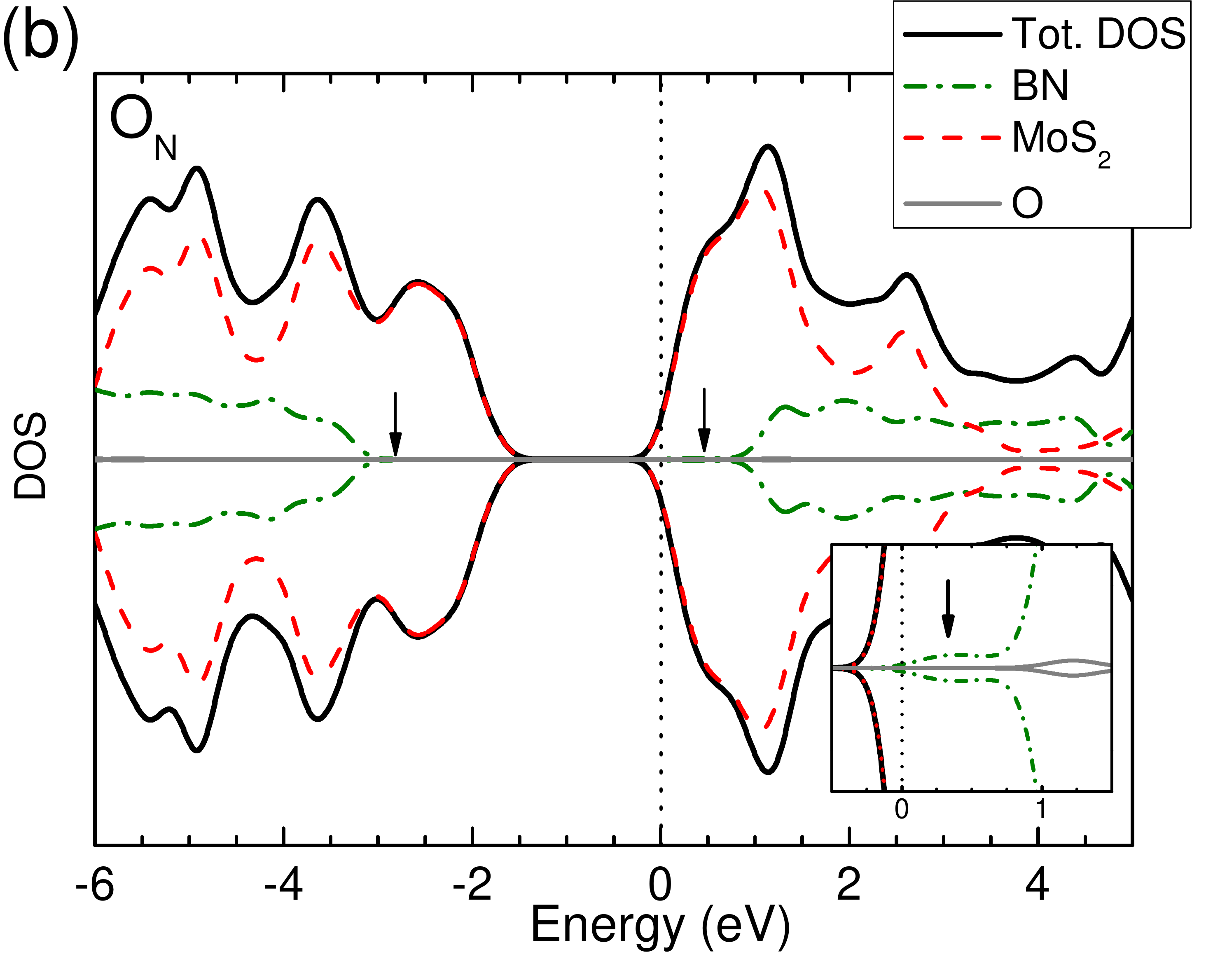}
\includegraphics*[width=0.28\textwidth]{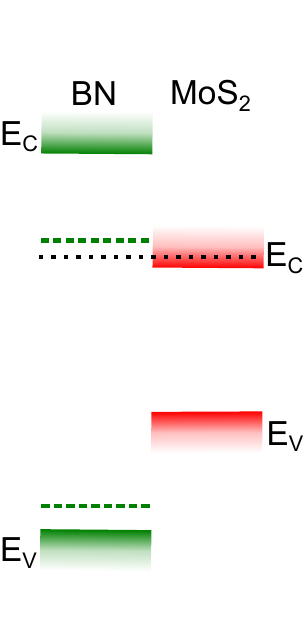}
\end{minipage}
\begin{minipage}{0.98\columnwidth}
\includegraphics*[width=0.95\textwidth]{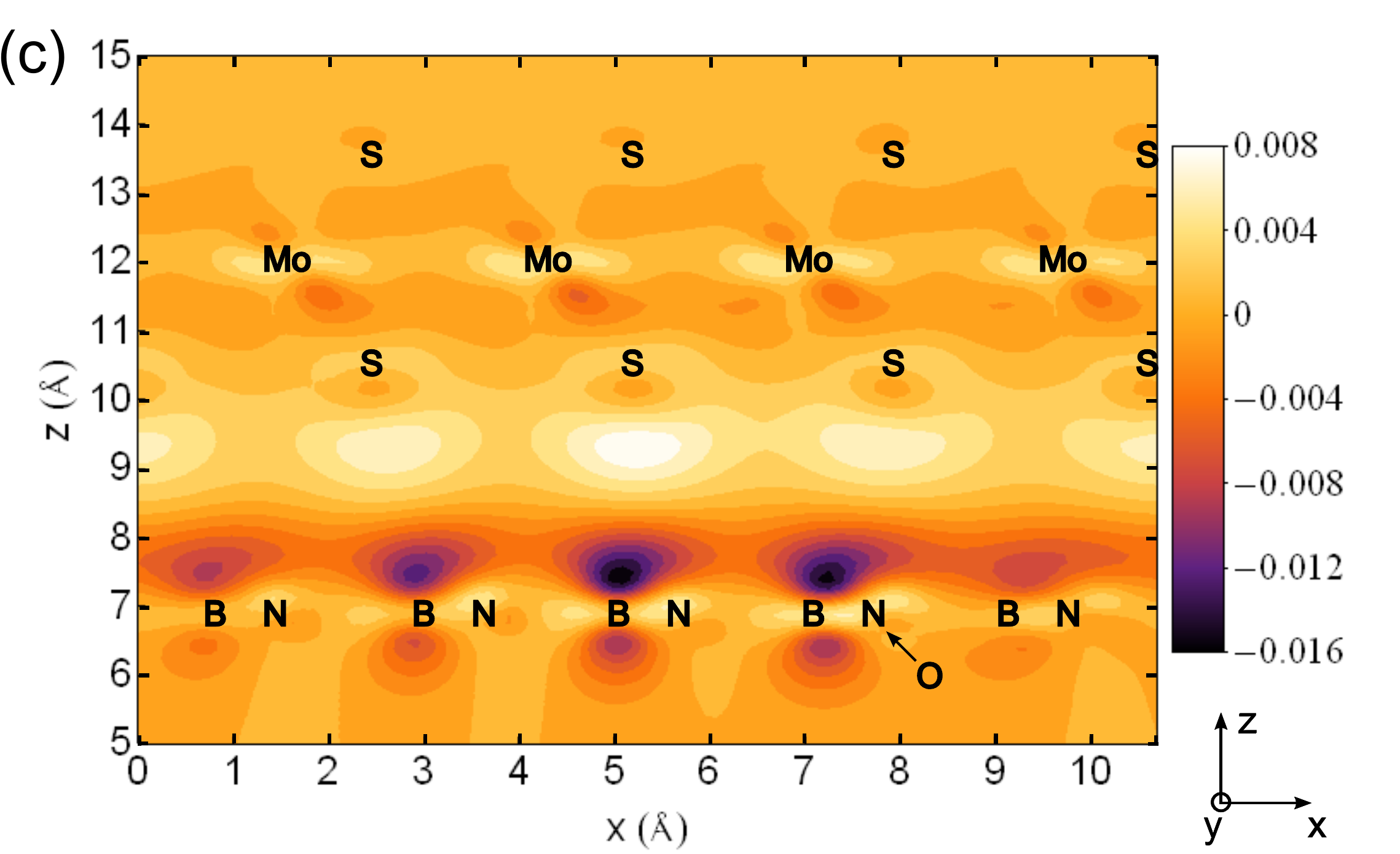}
\end{minipage}
\caption{\label{fig:n-dopants} (Color online) Calculated spin-resolved partial density of states plots from doping the BN subsystem with (a) $C_B$ and (b) $O_N$ substitutional defects. The zero-of-energy corresponds to the Fermi energy of the system. The dashed (green) lines in the band schemes show the relative positions of defect induced states in the BN band gap. (c) The difference of the charge densities (in electrons/\AA$^3$) of the oxygen-doped heterostructure and the isolated (oxygen-doped) BN and MoS$_2$ layers, averaged over the $y$-axis. The arrow shows the position of the oxygen impurity in the $x$-$z$-plane.
}
\end{figure}
A common form of \textit{non-native} defects are foreign atoms in the crystal lattice, which are either involuntarily incorporated into the lattice during growth as impurity atoms or intentionally as sources for electronic doping of the material. In order to assess the effect of $n$-doping of the BN layer on the electronic structure of the MoS$_2$ layer, we thus introduced C$_B$, O$_N$ and S$_N$ defects into the BN subsystem of the heterostructure. The carbon doping leads to a significant change of the geometry around the impurity, as the C-N bond lengths of 1.375\,\AA\space are much shorter than the B-N bond length of 1.44\,\AA\space in the defect-free layer due to the increased electron negativity of carbon compared to boron. For the same reason, the oxygen and sulfur impurities have the opposite effect on the defect geometry, pushing the surrounding boron atoms away. In case of the sulfur impurity, the difference in bond lengths of B-N (1.42\,\AA) and S-B bonds (1.84\,\AA) is such that the sulfur atom is pushed out of the BN plane into the interstitial region. The geometry in the MoS$_2$ layer, however, is not affected by the doping.

The changes in the electronic structure from carbon doping are similar to the case of the nitrogen vacancy, see Fig.~\ref{fig:n-dopants}~(a). In the absence of a MoS$_2$ layer, the impurity leads to a spin-polarized defect state of mixed nitrogen and carbon $2p$ character in the band gap, where the up-spin band is induced close to the conduction band edge of boron nitride and occupied with a single excess electron from the carbon impurity, which could be easily excited to the conduction band. The corresponding, unoccupied, down-spin band is resonant to the conduction band minimum\cite{park-2012,berseneva-2013}. With the MoS$_2$ layer present, the electronic bands of the MoS$_2$ realign compared to the defect free case and the BN bands, such that the energy of the defect state overlaps with the MoS$_2$ conduction band bottom. It is interesting that the presence of the MoS$_2$ and the corresponding charge transfer leads to a breakdown of the spin-polarization in the BN layer due to spin-compensation between the up- and down-spin states, which become degenerate. The observed realignment of the bandstructures of the two subsystems also seems to be independent of changes in the layer distance due to inclusion of van-der-Waals corrections. Correspondingly, a significant part of the excess carbon electron is transferred over to the MoS$_2$ layer, causing a small $n$-doping, similar to the case of the charge neutral nitrogen vacancy, and the spin-polarization of the carbon donator level is broken. We found a similar behaviour for oxygen, see Fig.~\ref{fig:n-dopants}~(b), and sulfur doping of the BN layer. The Fermi level pinning to the conduction band of MoS$_2$ and the corresponding realignment of the bands of the MoS$_2$ layer compared to those of BN lead to a significant change of valence band and conduction band offsets. Within the frame of the DFT-GGA functional, the valence band offset increases from 0.05\,eV in the defect-free case to a value of about 1.6\,eV, which leaves the conduction band offset at about the same size. We expect these offsets to be slightly larger under experimental conditions due to the intrinsic underestimation of electronic band gaps from DFT-GGA calculations. Doped BN thus remains a suitable substrate und tunnel material for $n$-type MoS$_2$. 

In order to obtain a better understanding of the nature of the charge transfer, we performed a projection on the $x$-$z$ plane of the difference of the electronic charge density of the heterostructure and the single (doped) BN and MoS$_2$ sheets, similar to what is shown for the defect-free heterostructure in Sec.~\ref{sec:pure}, Fig.~\ref{fig:pure}~(a). Fig.~\ref{fig:n-dopants}~(c) shows the plot for the heterostructure containing an oxygen impurity, which is representative for the other $n$-type dopants as well. As for the defect-free heterostructure, the plot reveals an accumulation of charge in the interstitial region that mainly stems from the BN layer. However, the amount of accumulated charge is an order of magnitude higher than for the pure heterostructure due to the additional and weakly bound charge from the oxygen impurity. The observed $n$-doping of the MoS$_2$ layer thus comes from partial occupation of antibonding sulfur $p_{z}$-orbitals at the bottom of the layer, which protrude into the interstitial region. This is further confirmed by an additional Mulliken charge analysis, which shows an general accumulation of electron density in the MoS$_2$ layer. The observed accumulation manifests in a twofold increase of the overall charge density on the sulfur atoms at the bottom from a value of $-0.05$\,\textit{e} to $-0.10$\,\textit{e}, while the density at the top sulfur atoms and the molybdenum atoms shows a light increase to a value of $-0.06$\,\textit{e} and $+0.12$\,\textit{e} (from $+0.13$\,\textit{e}), respectively. In contrast, the corresponding additional charge density from $n$-doping in the BN layer is mainly localized on the boron atoms adjacent to the defect, while the nitrogen atoms are unaffected.

\subsection{$p$-doping of BN}
\begin{figure}[tbh]
\begin{minipage}{0.98\columnwidth}
\includegraphics*[width=0.7\textwidth]{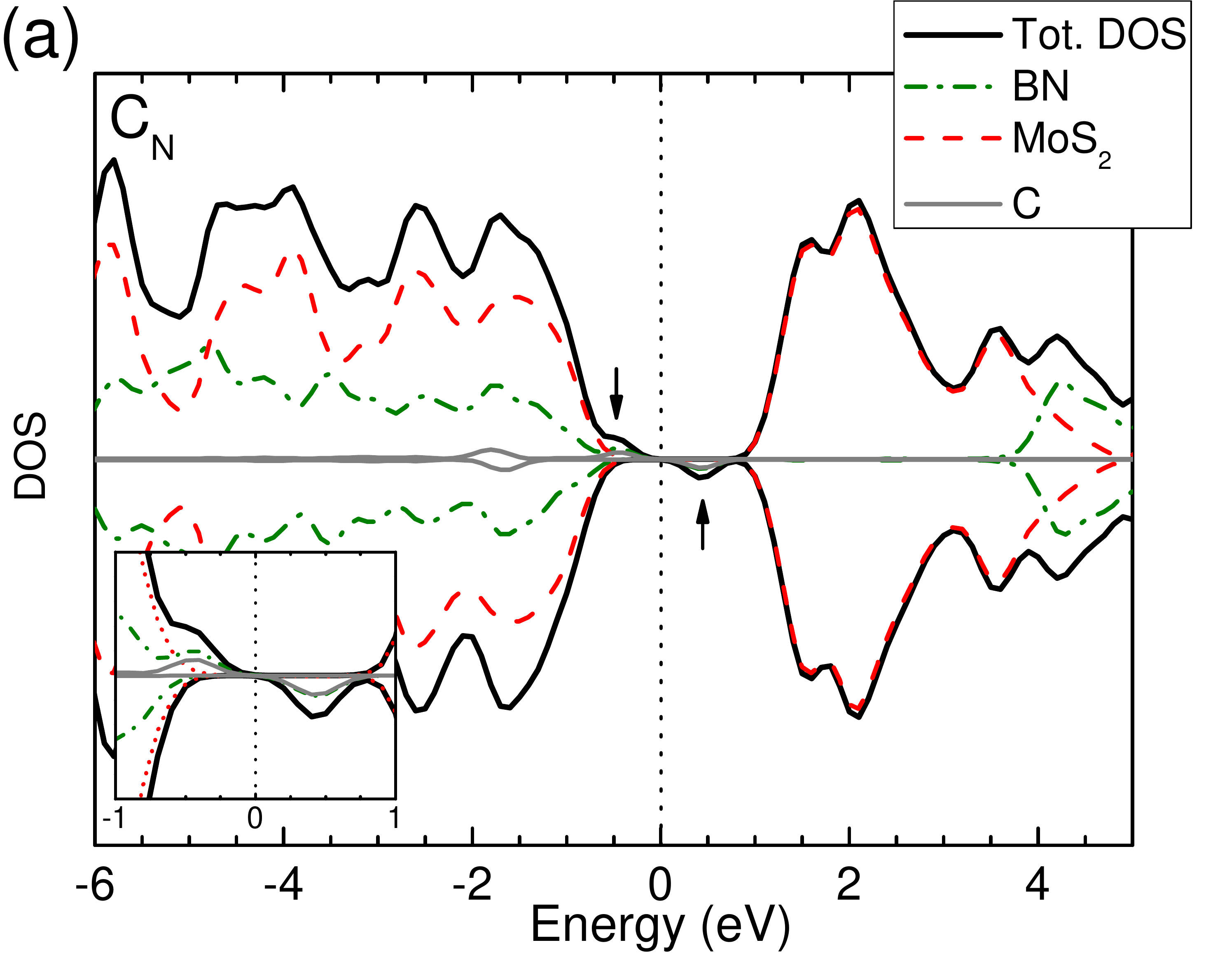}
\includegraphics*[width=0.28\textwidth]{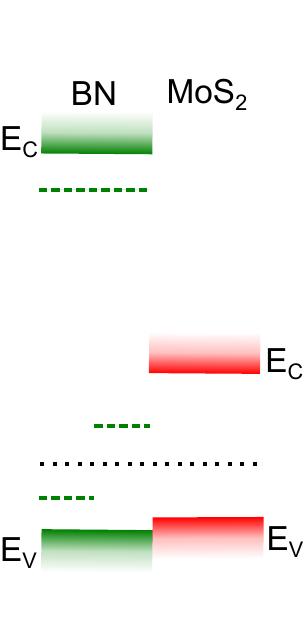}
\end{minipage}
\begin{minipage}{0.98\columnwidth}
\includegraphics*[width=0.7\textwidth]{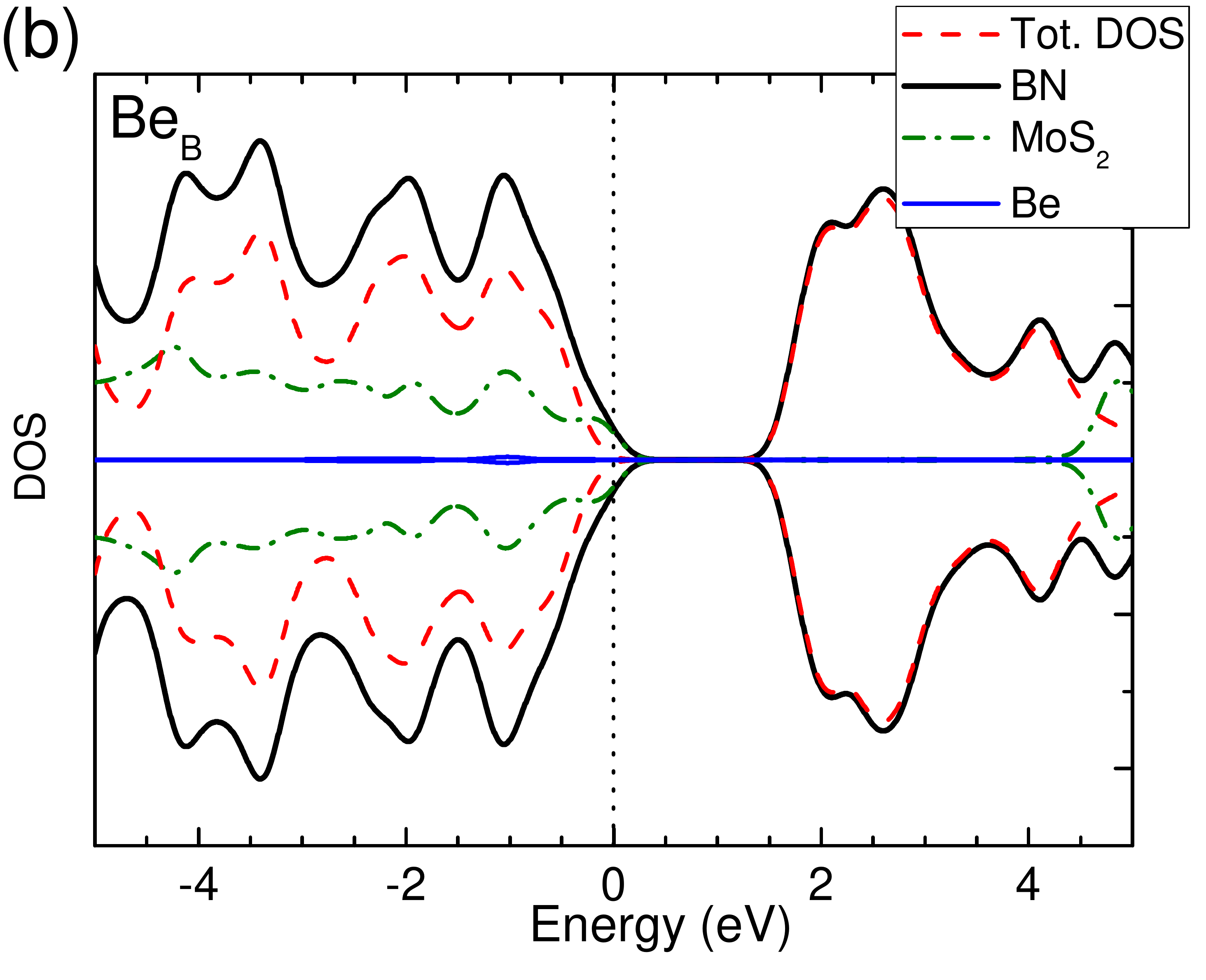}
\includegraphics*[width=0.28\textwidth]{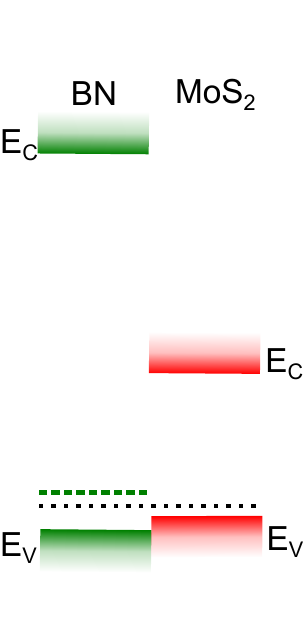}
\end{minipage}
\begin{minipage}{0.98\columnwidth}
\includegraphics*[width=0.95\textwidth]{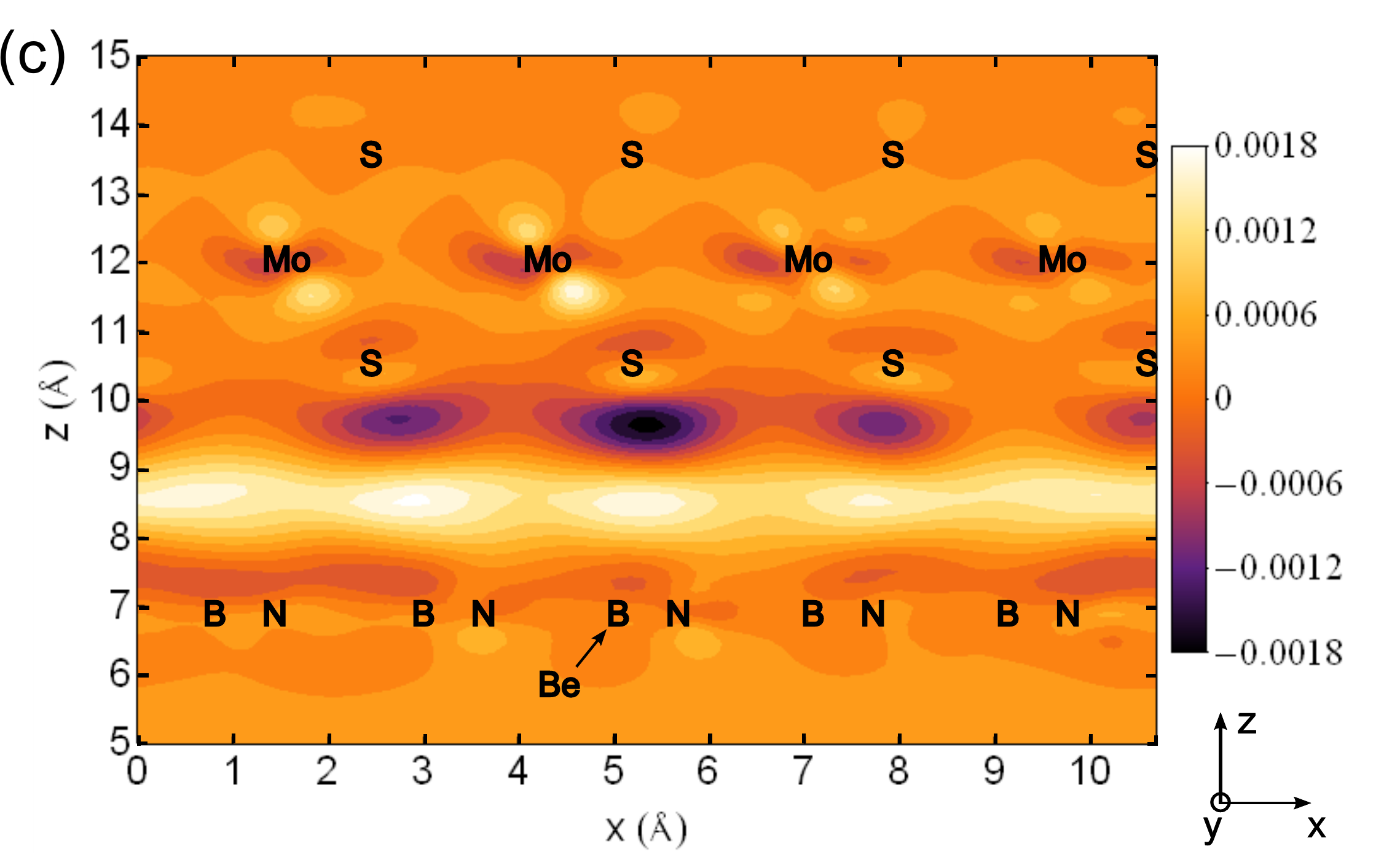}
\end{minipage}
\caption{\label{fig:p-dopants} (Color online) The calculated spin-resolved partial density of states plots for (a) C$_N$ and (b) Be$_B$ $p$-type substitutional defects in the MoS$_2$/BN heterostructure. The dotted (black) and the dashed (green) lines denote the Fermi energy and the defect-levels, respectively. (c) Same as Fig.~\ref{fig:n-dopants}~(c) for beryllium-doping.
}
\end{figure}
In the next step, we studied the effects on MoS$_2$ from $p$-doping of the BN layer by C$_N$ and Be$_B$ defects. Contrary to $n$-dopants, the $p$-dopants studied here all cause a breathing-like relaxation of the first atomic shell away from the impurity. In case of the C$_N$ defect, the bond lengths between carbon and boron increase compared to the B-N bond by about 3.8\%, from 1.445\,\AA\space to a value of 1.5\,\AA. Correspondingly, the B-N bond lengths in the second shell decrease slightly by 0.5\% to 1.438\,\AA. The geometry of the Be$_B$ defect is very similar to that of the boron vacancy. The nitrogen atoms surrounding the vacancy experience a breathing relaxation by 2.8\%, which decreases the bond lengths to 1.404\,\AA, very close to the value obtained for the boron vacancy. Correspondingly, the Be-N bond lengths are quite long, about 1.55\,\AA. In both cases, the geometry in the MoS$_2$ layer is again unaffected by the impurities. 

While the previously described $n$-type carbon impurity changes the electronic structure in a similar way as the nitrogen vacancy, the effect from a carbon replacing a nitrogen atom (C$_N$) is quite similar to that of the boron vacancy. As can be seen from the PDOS plot in Fig.~\ref{fig:p-dopants}~(a), the C$_N$ defect induces an occupied up-spin band of boron $2p$ character with some contribution from carbon just above the valence band maxima of BN and MoS$_2$. Another, unoccupied, spin-down band of complementary boron and carbon contributions is located in the middle of the MoS$_2$ band gap, in good agreement with recent DFT studies on C$_N$ defects in pure hexagonal boron nitride\cite{park-2012,berseneva-2013}. In contrast to the case of $n$-dopants, the bandstructure of the MoS$_2$ layer thus is essentially unaffected by the presence of the carbon impurity and the valence band offset between BN and MoS$_2$ remains small. 

The Be$_B$ defect induces one spin-degenerate defect level of mainly nitrogen character slightly above the valence band of BN, which is partly occupied with one electron, see Fig.~\ref{fig:p-dopants}~(b). The Fermi energy thus crosses the valence band maximum of the defective BN layer, and causes $p$-type conditions. The electronic bands of the MoS$_2$ layer are slightly shifted to lower energies and do not seem to be affected by the Be impurity. As for boron vacancy and the C$_N$ defect, there is no large change in valence and conduction band offsets found for the $n$-dopants due the weak realignment of the electronic structures of the two-subsystems originating from negligible inter-layer charge transfer in this configuration. 

On the other hand, an analysis of the $y$-averaged charge density in Fig.~\ref{fig:p-dopants}~(c) draws a slightly different picture. While there is no significant change of the accumulated charge in the interstitial region between the layers compared to the pure heterostructure, we find a noticeable transfer of charge from the MoS$_2$ layer directly above the Be atom. Indeed, we find a corresponding decrease of the Mulliken population of the sulfur atom directly above the Be impurity from $-0.05$\,\textit{e} in the defect-free heterostructure to $-0.04$\,\textit{e}. The other atoms in the MoS$_2$ seem to be not affected. In the BN layer, the effect of the impurity manifests in a significant charge transfer from the boron atoms in the second shell to the nitrogen atoms adjacent to the Be impurity. The corresponding Mulliken charges change from $+0.82$\,\textit{e} and $-0.83$\,\textit{e} in the defect-free structure to $+0.76$\,\textit{e} and $-0.91$\,\textit{e} for the boron and nitrogen atoms, respectively. We conclude from these findings that compared to the case of $n$-dopants, the effect of $p$-type impurities on the electronic structure of the MoS$_2$ and the charge density in the interstitial region appears to be small.

\subsection{Intercalated Sodium}
\begin{figure}[tbh]
\begin{minipage}{0.98\columnwidth}
\includegraphics*[width=0.7\textwidth]{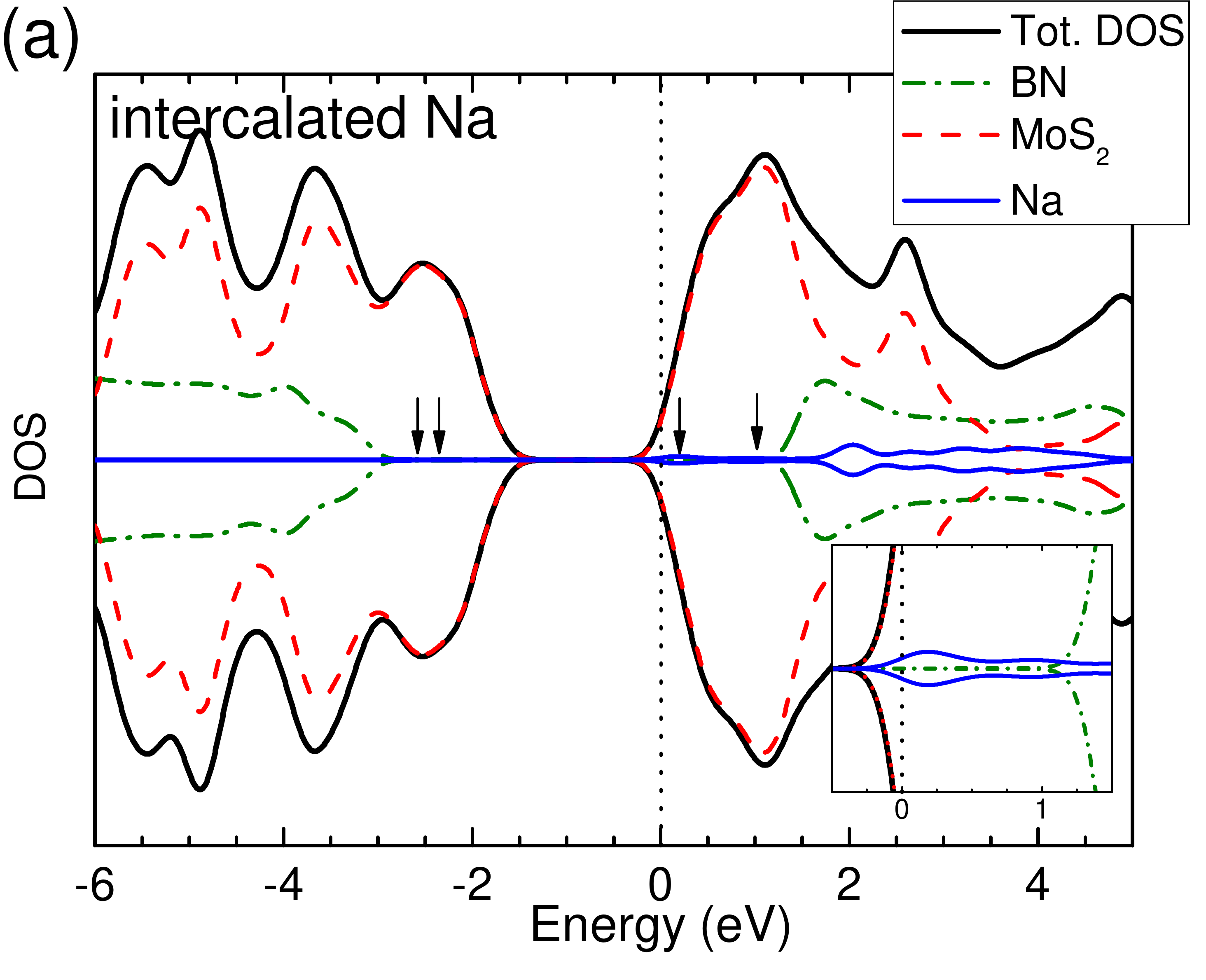}
\includegraphics*[width=0.28\textwidth]{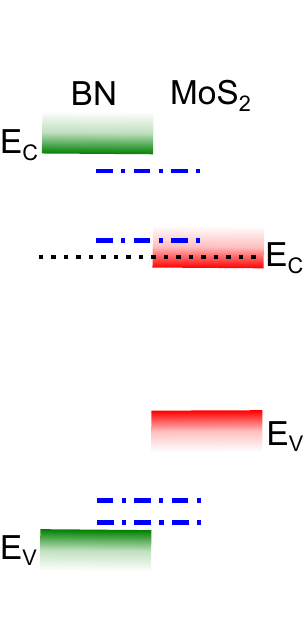}
\end{minipage}
\\
\begin{minipage}{0.98\columnwidth}
\includegraphics*[width=0.95\textwidth]{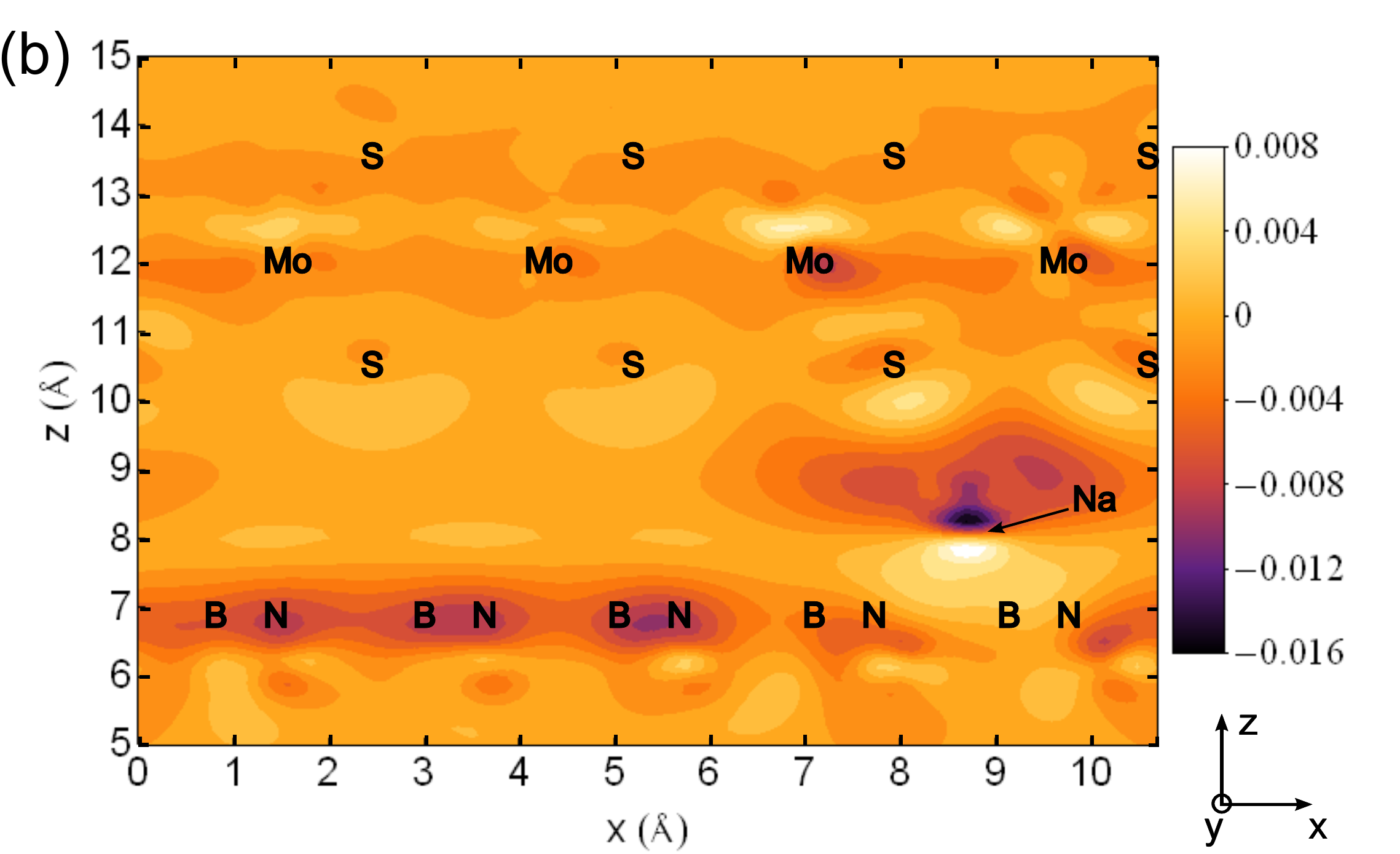}
\end{minipage}
\caption{\label{fig:intercal} (Color online) (a) Spin-resolved partial density of states and (b) the $y$-averaged charge density difference for a MoS$_2$/BN heterostructure containing a single Na atom between the layers. The charge density difference was calculated by substracting the charge densities of the single BN and MoS$_2$ sheets and an isolated Na atom from that of the compound heterostructure. As before, the zero-of-energy corresponds to the Fermi energy of the system. The dash-dotted (blue) lines in the right subfigure in (a) denote the relative energies of Na states in the band scheme.
}
\end{figure}
It is well-known that atoms intercalated between interfaces can have significant effects on the electronic properties of heterostructures. Dolui \textit{et al.}\cite{dolui-2013} recently predicted from DFT calculations that the experimentally found intrinsic $n$-doping of MoS$_2$ on SiO$_2$ could be explained by the existence of potassium atoms in the interfacial region between MoS$_2$ and the substrate, which are possible residues from chemical exfoliation based methods for the preparation of single or few-layer MoS$_2$ samples. 

Due to the likely occurrence of such intercalated impurities in experimentally prepared heterostructures, we simulated the effect of single intercalated sodium atoms on the electronic bandstructures of heterostructures of MoS$_2$ and BN sheets. We tested several initial configurations of the Na impurity with respect to the MoS$_2$ layer and optimized the geometry. In the final configuration with the lowest total energy, the Na atom assumes a position at about 1/3 of the interlayer distance above the BN layer, in the center of a MoS$_2$ hexagon. The geometries of the layers are not altered, except for an increase of the distance between the BN and the Mo layers from 5.1\,\AA\space to a value of 6.3\,\AA.

With regard to the electronic structure, intercalated Na atoms introduce a number of additional states into the band structure of the compound system. A partly occupied state with a weak dispersion appears about 1\,eV below the conduction band minimum of BN, while a band of unoccupied states is in resonance with the BN conduction band, see Fig.~\ref{fig:intercal}~(a). Similar to what we found for the nitrogen vacancy and $n$-dopants, the MoS$_2$ bands realign and move to lower energies relative to the BN bands such that the MoS$_2$ conduction band edge overlaps with the partially occupied Na state, and the heterostructure  assumes type-I nature with $\Delta E_V>0$ and $\Delta E_C<0$ when going from the BN to the MoS$_2$ layer. At the same time, the realignment of the electronic bandstructure causes the Fermi energy to cross the conduction band edge of the MoS$_2$ subsystem, populating the conduction band at the K valley. 

Intercalation of sodium into the system thus leads to a metallic MoS$_2$ layer, reminescent of what was predicted before\cite{andersen-2012} for bulk MoS$_2$ with intercalated potassium, and similar to the effect of $n$-doping the boron nitride layer. However, closer examination of the charge density shows some interesting differences. The distribution of the increased electron density over the interstitial region, and thus the sulfur atoms on the face pointing towards the BN layer and the Mo atoms, is nearly homogenous for the cases of $n$-doped BN. In contrast, the charge transfer from intercalated sodium mainly affects the sulfur (and molybdenum) atoms close to the impurity. The effect on the other atoms clearly decays with increasing distance, as is nicely seen in the $y$-averaged density difference plot in Fig.~\ref{fig:intercal}~(b). The Mulliken population analysis shows a significant change of the charge of the bottom sulfur and molybdenum atoms making up the hexagon closest to the intercalated Na from $-0.05$\,\textit{e} to $-0.16$\,\textit{e} and from $+0.11$\,\textit{e} to $+0.07$\,\textit{e}, respectively, while the increase at the other sulfur atoms at the bottom face is more moderate, from $-0.05$\,\textit{e} to $-0.08$\,\textit{e}. Similarly, we find an inhomogeneous change in the electron density in the BN layer as well. The charge density difference plot in Fig.~\ref{fig:intercal}~(b) suggests that the charge density in the region between the Na atom and the BN layer increases, which is caused by redistribution of the electronic charge from the Na atom and the boron atoms.

\section{Conclusion}
In conclusion, our study suggests that doping of the boron nitride layer in a bilayer MoS$_2$/BN heterostructure can have noticeable effects on the electronic properties of the MoS$_2$ layer. While $n$-doping of the boron nitride subsystem and the presence of nitrogen vacancies cause charge transfer from BN to the MoS$_2$ layer, $p$-type doping by C$_N$ and Be$_B$ substitutional defects has only minor effects. In agreement with recent findings on SiO$_2$ substrates, our study suggests that intercalated alkali metal atoms (in our case sodium) also cause $n$-type conditions in the MoS$_2$ subsystem, hereby populating the antibonding S 2$p_z$ states protruding into the interstitial region between BN and MoS$_2$ layers. Doping of BN substrates thus forms a possible cause for unintentional $n$-doping or local doping compensation in deposited MoS$_2$ layers. In the same way, it may be used intentionally for indirect $n$-doping while leaving the structural properties in the MoS$_2$ sheet unaffected. 

\section{Acknowledgements}
The authors gratefully acknowledge the Cambridge High Performance Cluster and the North-German Supercomputing Alliance (HLRN) for providing the computational ressources used for the simulations in this work. This work was supported in part by the European Research Council (ERC) under grant number 259286.

\bibliographystyle{apsrev4-1}

\end{document}